\documentclass[10pt,prd,twocolumn,showpacs,amssymb,superscriptaddress,aps,nofootinbib,noeprint,nolongbibliography
]{revtex4-2}

\usepackage{graphicx,amsmath,amssymb}
\usepackage{bm}
\usepackage[colorlinks=true]{hyperref}
\hypersetup{colorlinks=true,
            citecolor=blue,
            linkcolor=red,
            urlcolor=red}

\usepackage{amssymb,amsmath,amsfonts,mathrsfs,url}
\usepackage{xcolor}
\usepackage[colorlinks=true]{hyperref}
\hypersetup{citecolor=blue,linkcolor=red}
\usepackage{multirow,array}
\DeclareMathAlphabet{\pazocal}{OMS}{zplm}{m}{n}

\usepackage{pifont}

\newcommand{\Ms}{M}
\newcommand{\Rs}{R}

\def\apj{{ApJ}}

\def\apjl{{ApJL}}

\def\aap{{A\&A}}

\def\mnras{{MNRAS}}

\def\nat{{Nature}}

\def\prd{{Physical Review D}}
\def\prl{{Phys. Rev. Lett.}}

\def\04a{{2004 a}}
\def\04b{{2004 b}}

\begin{document}

\title{Universality of gravitational radiation from magnetar magnetospheres}
\date{\today}
\author{Arthur G. Suvorov}\thanks{arthur.suvorov@tat.uni-tuebingen.de}
\affiliation{Departament de F{\'i}sica, Universitat d'Alacant, Ap. Correus 99, E-03080 Alacant, Spain}
\affiliation{Theoretical Astrophysics, IAAT, University of T{\"u}bingen, T{\"u}bingen, D-72076, Germany}
\author{Petros Stefanou}\thanks{petros.stefanou@ua.es}
\affiliation{Departament de F{\'i}sica, Universitat d'Alacant, Ap. Correus 99, E-03080 Alacant, Spain}
\author{Jos{\'e} A. Pons}\thanks{jose.pons@ua.es}
\affiliation{Departament de F{\'i}sica, Universitat d'Alacant, Ap. Correus 99, E-03080 Alacant, Spain}

\begin{abstract}
\noindent The intense magnetic fields inferred from magnetars suggest they may be strong gravitational-wave emitters. Although emissions due to hydromagnetic deformations are more promising from a detection standpoint, exterior fields also contribute a strain. However, numerical evidence suggests that the free energy of stable magnetospheric solutions cannot exceed a few tens of percent relative to the potential state, implying that the magnetospheric contribution to the gravitational-wave luminosity cannot differ significantly between models. This prompts `universality', in the sense that the strain provides a direct probe of the near-surface field without being muddied by magnetospheric currents. Using a suite of three-dimensional, force-free, general-relativistic solutions for dipole and dipole-plus-quadrupole fields, we find that space-based interferometers may enable marginal detections out to $\lesssim$~kpc distances for slowly-rotating magnetars with fields of $\gtrsim 10^{15}$~G independently of internal deformations.
\end{abstract}

\maketitle

\section{Introduction} \label{sec:intro}

Magnetars are a class of neutron star that categorically boast strong magnetic fields ($\gtrsim 10^{14}$~G). Because of their strong fields, a sizeable mass-quadrupole moment may be induced in the interior through hydromagnetic deformations \cite{gg18}. As such, provided the magnetic and rotation axes are misaligned, young magnetars may be significant sources of gravitational waves (GWs). Models are, however, notoriously sensitive to the geometric structure of the internal field: depending on the poloidal-toroidal partition \citep{mast11,cr13,lm13}, multipolarity \citep{col08,c09,mast15}, equation of state \citep{hask08,fr12,dex19}, and the possible presence of superconducting protons \citep{lan13,sed25}, strains spanning some orders of magnitude have been predicted for a given surface-field strength.

Another channel for GW emission comes from the external field directly \citep{mir03,naz20,cont24,kouv24}. Although weaker than many predictions for internally-sourced contributions, the absence of fluid degrees of freedom means that these signals would carry different information and effectively set a floor value to the GW luminosity. Importantly, magnetospheric solutions are typically classified by their \emph{twist} relative to the potential (i.e. current-free) state. These currents, sourced via the internal-to-external transfer of helicity and crustal-platelet slippage, are often invoked to explain magnetar activity. For example, hotspot phenomenology may be tied to toroidal field evolution \cite{gepv14,mpm15}, emission spectra and Alfv{\'e}n wave transport are mediated by current bundles \cite{belo09,parfey13}, and radio activity is likely regulated by near-surface fluxes \cite{coop24,suvm23} as are overstraining `failures' in the outer crust \cite{perna11,pons11}. 

Despite the diversity of phenomena described above we show, by building a sequence of three-dimensional (3D) and general-relativistic (GR) twisted magnetospheres following \citet{stef25} (henceforth SSP25), that the external GW signal is \emph{universal}. Because there is a maximum excess energy, relative to an untwisted state, that the magnetosphere can sustain before the (elliptic) force-free problem no longer admits (stable) solutions \citep{bin72,aly84,mahl19}, we argue that strains cannot vary much between models. The magnitude of magnetospheric GW emissions is therefore set mainly by the near-surface field and stellar radius in some combination, while the polarisation phase portrait is set by the distribution of surface currents defining the boundary conditions \cite{cont24,naz20,kouv24}. A future detection could thus provide a probe of these quantities, which are especially difficult to infer in high-B objects due to their intrinsically noisy nature \cite{mel99,suvm23,low25}. 
{We focus on magnetospheric signals in this paper, ignoring the fluid deformation (which may be larger).}

While the strains we anticipate have little chance to be detected with the currently-operational GW network, they may be visible to next-generation instruments like the Deci-hertz Interferometer Gravitational wave Observatory (DECIGO) most sensitive in the $\lesssim 10$~Hz band \cite{kaw21,pag25}. We quantify the signal-to-noise ratios (SNRs) expected for realistic integration windows for a variety of magnetospheric configurations, showing that marginal detections out to nearby ($\lesssim$~kpc) star-forming clusters like NGC 7822 are feasible in the next decade(s) for fields with strong multipolar components. 

This paper is organised as follows. Section~\ref{sec:magnetospherebasics} briefly recaps the theory of force-free exteriors, together with our numerical methods (Sec.~\ref{sec:numerics}) and why the magnetosphere can sustain a maximum twist (Sec.~\ref{sec:maxtwist}). GW strains are introduced in Section~\ref{sec:gws} and computed for a variety of models in Section~\ref{sec:magmodels}. Detection prospects are covered in Section~\ref{sec:detect} for dipolar (Sec.~\ref{sec:dipsnr}) and multipolar (Sec.~\ref{sec:dipquadsnr}) configurations. Some closing discussion is given in Section~\ref{sec:discussion}.

\section{Static magnetospheres} \label{sec:magnetospherebasics}

In this work, the primary equation of relevance describing
the magnetospheric structure is that describing (GR) force-free systems in the static limit\footnote{We note
that we work exclusively with the magnetic 3-vector, $\bm{B}$, defined by an orthonormal tetrad. Where convenient and no ambiguity can arise, we adopt units such that $G=c=1$.},
\begin{equation}\label{eq:magnetoeqn}
    \bm{\nabla} \times (e^{\nu}\bm{B}) = \alpha \bm{B},
\end{equation}
where $\alpha(\boldsymbol{x})$ effectively encapsulates the toroidal twist. This parameter has to obey the additional constraint 
\begin{equation}\label{eq:alpha_constraint}
    \bm{B} \cdot \bm{\nabla} \alpha = 0,
\end{equation}
meaning that it should be constant along field lines. In the above and throughout, we work with Schwarzschild coordinates such that $\bm{\nabla} = \{e^{-\lambda} {\partial_r} ,  r^{-1}{\partial_\theta}, + r^{-1} \csc \theta{\partial_\phi}\}$
denotes the 3-gradient operator associated with the line element
\begin{equation} \label{eq:sch}
    ds^2 = -e^{2\nu} dt^2 + e^{2\lambda} dr^2 + r^2d\theta^2 + r^2 \sin^2{\theta} d\phi^2.
\end{equation}
Outside of slow stars with $B \ll 10^{18}$~G, the metric potentials are well approximated by $e^{\nu} = e^{-\lambda} = \sqrt{1 - {2\Ms}/{r}}$ due to Birkhoff's theorem. It is convenient to introduce the compactness, $\mathcal{C} = \Ms / \Rs$, to categorise GR models for mass $\Ms$ and radius $\Rs$; a typical value for a neutron star is $\mathcal{C} \approx 0.17$ \cite{ofen24,skk24}.

By solving equations \eqref{eq:magnetoeqn} and \eqref{eq:alpha_constraint} for a variety of physically-motivated setups, we investigate GW predictions. In SSP25, these were solved using physics-informed neural networks (PINNs; see Sec.~\ref{sec:numerics}). In the untwisted ($\alpha = 0$) and axisymmetric limit, multipolar solutions to \eqref{eq:magnetoeqn} can be described by \cite{col08,c09}
\begin{equation}
\begin{aligned} \label{eq:puredip}
B_{r_0}=B_{0}\frac{ \psi_{,\theta}}{r^2\sin{\theta}}, \quad B_{\theta_0}=-B_{0}\frac{e^{-\lambda} \psi_{,r}}{r\sin{\theta}}, \quad B_{\phi_0} = 0,
\end{aligned}
\end{equation}
in terms of the poloidal flux
\begin{equation} \label{eq:psifn}
    \psi = \sin \theta \sum_{\ell} \frac{\mu_{\ell}} {r^{\ell}} ~F\left(\left[\ell,\ell+2\right], \left[2+2\ell\right], \frac{2 \Ms}{r}\right) \frac{d P_{\ell}(\cos \theta)}{d \theta},
\end{equation}

where $F$ is the extended hypergeometric function and $\mu_{\ell}$ together with $B_0$ set the contributions of multipoles and the overall field strength, respectively. Expression \eqref{eq:puredip} provides a useful reference point
with which twisted solutions can be compared. Although including an arbitrary number of multipoles within the PINN solver is not particularly difficult, we concentrate on the case of dipole-plus-quadrupole fields for simplicity. The Newtonian limit for this configuration is described (for instance) by equation (13) in \citet{mast15}, which is recovered in our notation for $\mu_{1} = -\Rs^3$ and $\mu_{2} = -\kappa \Rs^4/3$, where $\kappa$ sets the relative strength between the quadrupole and dipole components.

As above and throughout this paper, we refer to \emph{twist}. What exactly this means is, to some degree, subject to definitional variation (see Appendix C of \citet{liu16} for a survey). While different notions may be more appropriate for discussing a given aspect of magnetospheric problems (e.g., defining twist as the integral of the torsion along parametric curves for mathematical aspects of solution existence) it is sufficient, for our purposes, to refer to maximum values of $\alpha$ in the context of some given profile (see Sec.~\ref{sec:numerics}). This is because our primary interest concerns GW strains rather than magnetospheric aspects directly. In this context, we mean \emph{twist} to refer to a state where the global magnetic energy is greater than that of a potential field as $B_{\phi}\neq0$ because of the presence of currents. 

\subsection{Numerical methods} \label{sec:numerics}

SSP25 demonstrated how one may solve equations \eqref{eq:magnetoeqn} and \eqref{eq:alpha_constraint} using a PINN solver. Here we provide a brief recap of the numerical implementation in that work to which we refer the interested reader for a more in-depth description. 

In the PINN framework, the solution of the coupled system \eqref{eq:magnetoeqn} and \eqref{eq:alpha_constraint}, along with the solenoidal condition $\bm \nabla \cdot \bm B = 0$, is approximated by a fully-connected neural network $\bm{B}(\bm{x}; \bm{\Theta}), \alpha(\bm{x}; \bm{\Theta})$ whose parameters $\bm{\Theta}$ are optimized to minimize a loss function.
This loss is defined as the mean-squared residual of the PDEs evaluated at $N = 8 \times 10^{4}$ sampled points in a compactified domain and reflects how well the PINN approximation satisfies the equations. 
Training involves iteratively adjusting $\bm{\Theta}$ until the loss converges below a threshold or a maximum iteration limit is reached. Boundary conditions are enforced via a hard-enforcement strategy, meaning the network outputs are modified to inherently satisfy these conditions before being used in the loss computation.

The models studied in this work consist of dipole or dipole-plus-quadrupole fields for a pair of surface `hotspots' connected through a twisted magnetic loop. The radial magnetic field at the surface, $B_{r*},$ is defined by the potential solution \eqref{eq:puredip}. For $\alpha(\bm{x})$, we chose a Gaussian profile in the region where $B_{r*} > 0$ given by
\begin{equation} \label{eq:alphaexpression}
    \alpha_* = \alpha_0 \exp{\left[-\frac{(\theta - \theta_1)^2 +(\phi - \phi_1)^2}{2\sigma^2}\right]},
\end{equation}
where $\alpha_0$, $\theta_1$, and $\sigma$ are parameters that control the current strength, location, and size of the spot, respectively ($\phi_1$ is superfluous if only one hotspot pair is considered). This family sufficiently captures most of the phenomenological properties of magnetar hotspots (temperature, size, excess energy stored, ...) \cite{mpm15,youn22}.

\begin{figure}
    \includegraphics[width=0.5\textwidth]{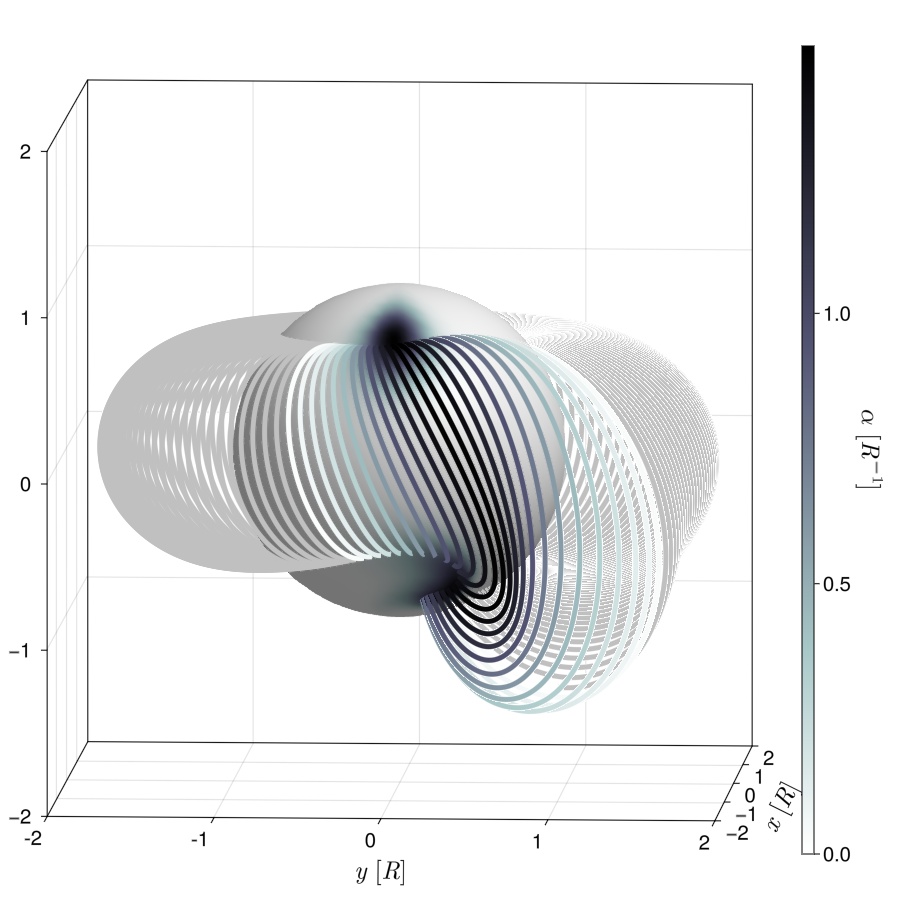} \\
    \includegraphics[width=0.5\textwidth]{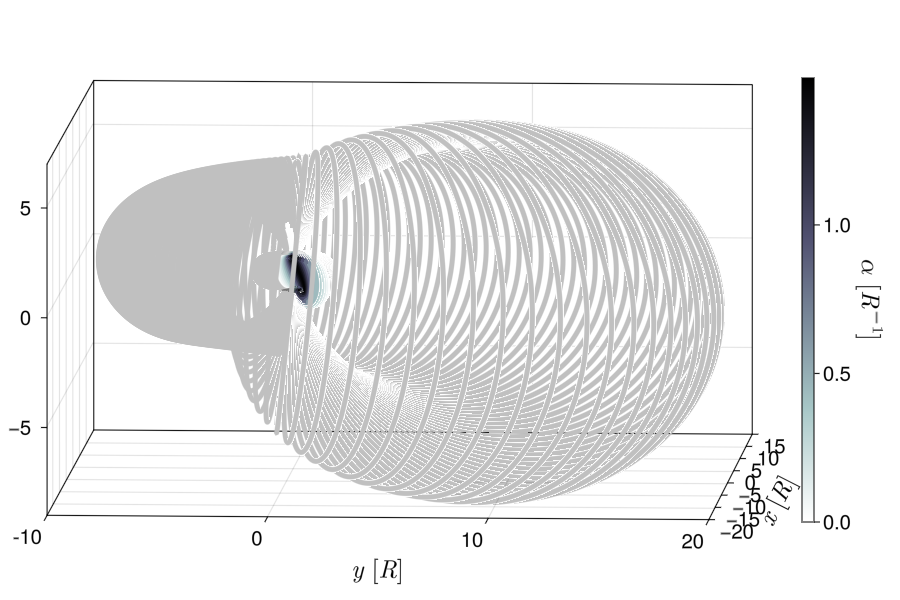}
    \caption{Magnetic field lines for $\alpha_{0} = 1.5 R^{-1}$, $\theta_{1} = \pi/4$, and $\sigma = 0.2$ near the stellar surface (top) and at larger radii (bottom) for a star with $\mathcal{C} = 0.17$. The colour bars show the position-dependent value of $\alpha(\bm{x})$ itself.
    \label{fig:dipolelines}}
\end{figure}

An example of field lines produced using PINNs for $\alpha_0 = 1.5 \Rs^{-1}, \theta_1 = \pi/4$, and $\sigma = 0.2$ is shown in Figure~\ref{fig:dipolelines} near the surface (top panel) and at large radii (bottom) for a canonical compactness $\mathcal{C} = 0.17$. Key features of the twist are clear: there is a toroidal pressure that inflates the magnetosphere in the right hemisphere relative to the left. The maximum value of $\alpha = 1.5 \Rs^{-1}$ is reached at the base of the `spots' defined by $\theta_{1}$, with relative cross-sectional width controlled by $\sigma$. These parameters correspond roughly to a pair of spots of size $\sim 2$~km, typical for magnetars (see figure 12 in \citet{belo16}). Note that the bottom-panel axes extend to $r \approx 15 \Rs$, highlighting the extent of toroidal blow-out.

\subsection{Maximum twists} \label{sec:maxtwist}

Numerous analytic \cite{chiu77,ein83} and numerical \cite{hp81,pai09} studies of equations \eqref{eq:magnetoeqn} and \eqref{eq:alpha_constraint} have identified that there is a maximum twist which the field can sustain in an infinite cavity until solutions no longer exist (see proposition II.3 in Ref.~\cite{aly84} for a proof for arbitrary star-shaped domains). In the classical limit of a plasma cylinder with longitudinal electric current, this maximal configuration is known as the Kruskal-Shafranov limit \citep{ks54}.

Aside from non-existence, multiple solutions for given boundary data can exist when $\alpha(\bm{x})$ is not constant, with some branches admitting total energies which can reach double that of the potential state (see, for instance, figure 8 in Ref.~\cite{flyer04}). \citet{koj17} found even larger excesses in axisymmetric, GR magnetospheres for very compact stars (factor $\sim$~few for $\mathcal{C}= 0.25$). These ultra-energised solutions are characterised by the presence of disconnected magnetic domains or `islands', seen in studies where the field is twisted via star-disk interactions \citep{uz02,matt05}, accretion-induced magnetic burial \citep{pm04,suvm19}, or crustal currents from magnetothermal evolutions \citep{akg18}. However, numerical evidence indicates that only the lowest-energy branch is stable (see Appendix C in SSP25), with plasmoids being ejected by kink or buoyant instabilities \citep{pm07,parfey13,mahl19}. Even if large excesses can exist mathematically, they are unlikely to be realised in Nature therefore: the free energy is limited to be of order $\approx 30\%$ in cases of axisymmetric twists -- which are always `global' in some sense -- or $\approx 5\%$ in realistic systems with localised twists \citep{stef23,stef25}. The inclusion of rotation does not seem to alleviate the problem as magnetic surfaces tend to spread radially unless arrested by hydrodynamic pressures a l{\'a} `magnetic towers' \cite{uz02}. Moreover, SSP25 found that the maximum energy excess of the stable branch is roughly constant as a function of compactness.

The fact that this maximum excess is small for stable configurations is at the heart of our universality results. As a harmonic-weighted magnetic energy-density alone features in the quadrupole moment integral for magnetospheric GW emissions (see Sec.~\ref{sec:gws}), it is not possible to exploit hydromagnetic freedoms to introduce large misalignments in the (magnetic) moments of inertia as for the interior problem \cite{gg18,mast11,cr13,lm13,col08,c09,mast15}. This means that GW signals anticipated from the magnetospheric fields of magnetars are entwined primarily with the surface field (see Sec.~\ref{sec:dipquad}). A future detection may therefore directly probe $B_{0}$, for which timing data are rather unreliable owing to the bumpy spindown of magnetars \cite{low25}.

\section{Gravitational radiation from magnetospheric fields} \label{sec:gws}

The characteristic GW luminosity of a source is typically quantified in terms of its gravitational multipole moments. While several studies have investigated the polarisation properties of magnetospheric signals \citep{naz20,cont24,kouv24}, we are here primarily interested in the amplitude itself for which knowledge of the strain, $h(t)$, is sufficient (though see also Sec.~\ref{sec:hydromag}). The best-suited detector for Galactic magnetars with spin frequencies $\lesssim 1$~Hz is DECIGO \citep{kaw21}. The design specifications of this space-based instrument indicates that the sensitivity bowl peaks at such frequencies. Detectability results are deferred to Sec.~\ref{sec:detect}.

\subsection{Wave strains and quadrupole moments} \label{sec:quadmoments}

The characteristic strain typically used in the GW literature reads \cite{aasi14}
\begin{equation} \label{eq:h0}
h_{0} = \frac{16 \pi^2 G}{c^4} \frac{|Q^{22}| f^2}{d},
\end{equation} 
for the leading-order ($\ell = m  = 2$) mass quadrupole moment $Q^{22}$, (linear) spin frequency $f$, source distance $d$, Newton's constant $G$, and speed of light $c$. The first of these quantities is typically that which is the most uncertain in astrophysical applications. In general, $Q^{22}$ has various contributions that appear at different post-Newtonian (PN) orders involving integrals over the components of the Landau-Lifshitz pseudotensor \citep{thorne80}. At Newtonian order, only the $00$-component of the $\mathcal{O}(c^{0})$ piece of the stress-energy enters (i.e., the mass density), which is often thought to dominate in the interior \cite{mak14,mak21}. Note that in realistic applications where $h_{0}$ is tracked, one can instead consider functions of time for the relevant variables, i.e., $f(t)$ and possibly $Q^{22}(t)$.

There is, however, a direct magnetic contribution that enters at 1PN due to the effective magnetic pressure \cite{mast15}. For a domain consisting of a magnetosphere extending from the stellar surface ($\Rs$) out to some (fiducial) light-cylinder $(R_{\rm LC})$, this contribution takes the general form\footnote{The quadrupole formula is formally derived assuming a bounded domain and thus its application is, strictly speaking, invalid if one takes $R_{\rm LC} \to \infty$ (see \citet{mir03}). Even for rapidly rotating objects with $R_{\rm LC} \sim 10 \Rs$ the correction is negligible though. Note that we also ignore electric field contributions.}
\begin{equation} \label{eq:massquadnaxi}
\begin{aligned}
Q^{22} =& \int^{R_{\rm LC}}_{\Rs} dr \int^{\pi}_{0} d\theta \int^{2\pi}_{0} d \phi \Bigg\{ \frac{B^2}{8 \pi c^2} r^4\sin\theta \\
&\times \left[ \cos^2 \phi \left( 1 - \cos^2\theta \right) - \cos^2\theta \right]\Bigg\},
\end{aligned}
\end{equation}
where the $c^2$ in the denominator attests to the PN nature of the contribution. Note that the magnetospheric mass density can be estimated from the \citet{gj69} value:
\begin{equation} \label{eq:gj}
    \rho_{\rm 0} \approx \frac{f |B|} {c} \frac{m_{e}} {e},
\end{equation}
for electron mass and charge $m_{e}$ and $e$, respectively. Expression \eqref{eq:gj} is $\sim 8$ orders of magnitude smaller than the PN contribution, $B^2 / 8 \pi c^2$, even for $B \lesssim 10^{12}$~G and $f \gtrsim 1$~Hz.

\begin{figure}
    \includegraphics[width=0.49\textwidth]{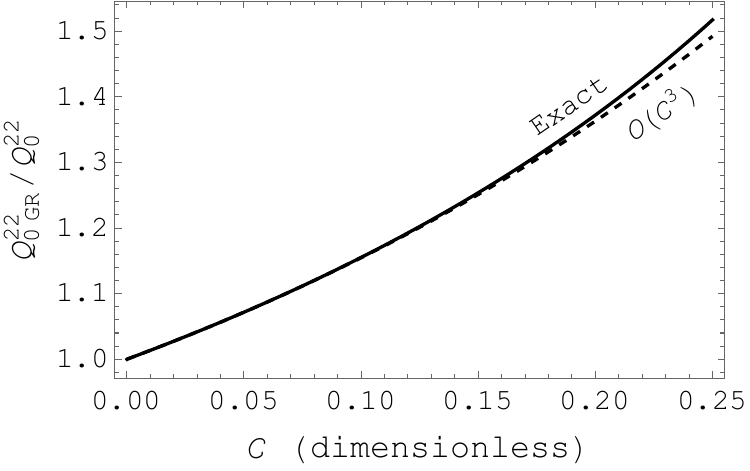}
    \caption{Comparison of the GR-boosted mass-quadrupole moment for an untwisted dipole relative to the Newtonian case, using either the perturbative expansion \eqref{eq:q22} (dashed) or direct integration (solid).
    \label{fig:normC}}
\end{figure}

Given a magnetospheric solution, expression \eqref{eq:massquadnaxi} can be evaluated to estimate the strain \eqref{eq:h0}. As a baseline model, one can calculate \eqref{eq:massquadnaxi} analytically for the untwisted dipole ($\kappa = 0$). This forms a reference value which we call $Q^{22}_{0}$ to compare with in subsequent sections. The result is
\begin{equation} \label{eq:q22}
\begin{aligned}
Q^{22}_{0} &\approx    -\frac{B_{0}^2 \Rs^5}{5 c^2} \left(1+\frac{4 \mathcal{C}}{3} + \frac{28 \mathcal{C}^2}{15} + \frac{27 \mathcal{C}^3}{10}\right) + \mathcal{O}(\mathcal{C}^4), \\
&\approx -2.2 \times 10^{38} B_{15}^2 R_{6}^5 \left(1+0.28 M_{1.4} R_{6}^{-1} \right) \text{ g cm}^{2},
\end{aligned}
\end{equation}
where the first approximation assumes $R_{\rm LC}/\Rs \gg 1$ -- as expected for a slow magnetar -- and we have introduced the (CGS units) normalisations $B_{15} = B_{0}/(10^{15}\text{ G})$, $M_{1.4} = \Ms/1.4 M_{\odot}$, and $R_{6} = \Rs/(10^{6} \text{ cm})$. Curiously, we see that the magnetospheric $Q^{22}$ is \emph{negative} for poloidal fields, reflecting the fact that $B^2$ scales with negative powers of radius rather than positive ones in the interior (as demanded by regularity towards the origin). 

\begin{figure*}
    \includegraphics[width=0.32\textwidth]{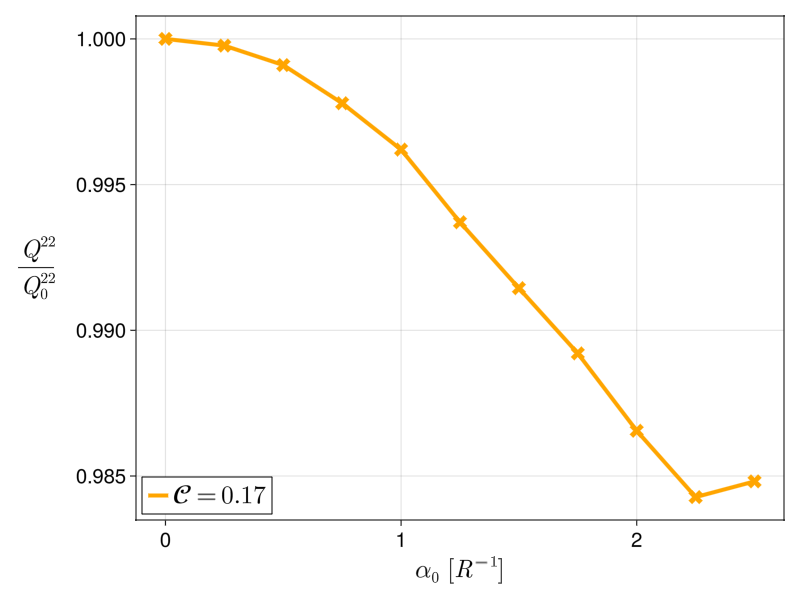}
    \includegraphics[width=0.32\textwidth]{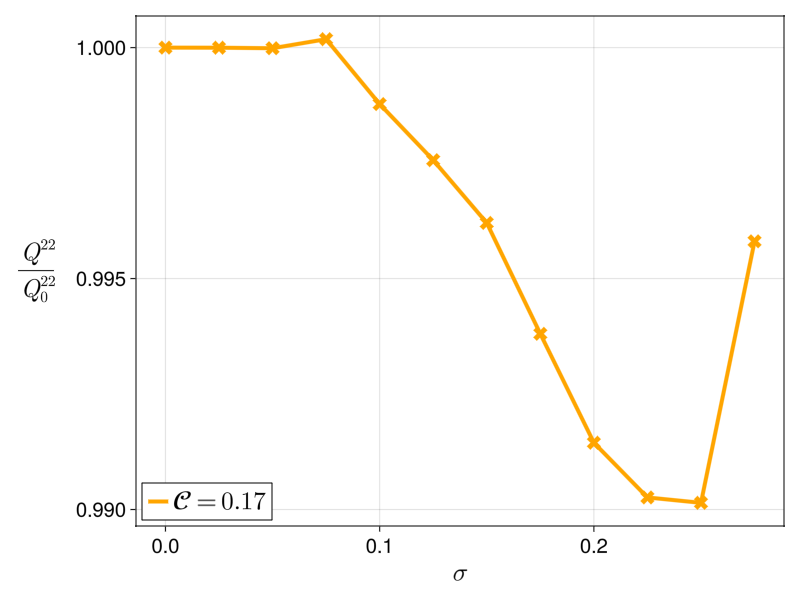}
    \includegraphics[width=0.32\textwidth]{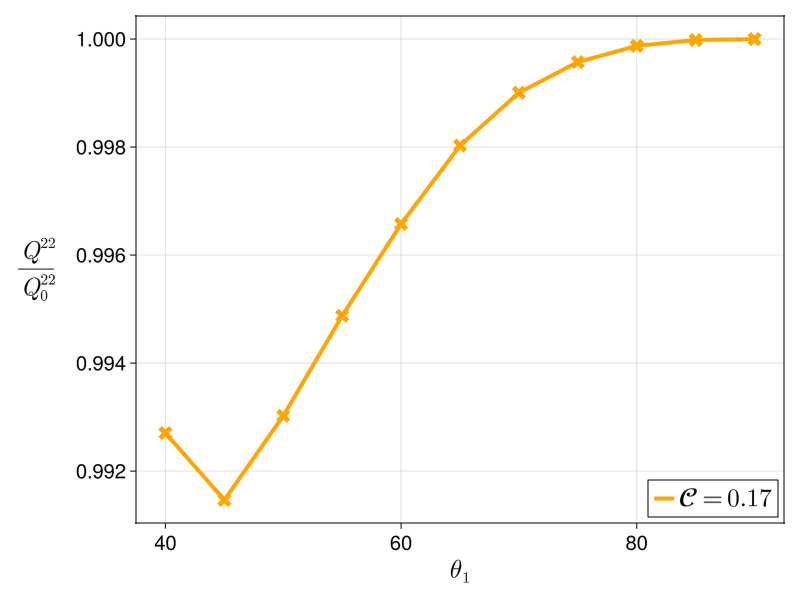}
    \caption{Magnetospheric quadrupole moments \eqref{eq:massquadnaxi} for compactness $\mathcal{C} = 0.17$, normalised to the untwisted dipole, for varying base twist ($\alpha_{0}$; left), spot scale ($\sigma_{0}$; middle), and coronal-loop latitude ($\theta_{1}$ in degrees; right) as defined by expression \eqref{eq:alphaexpression}. 
    \label{fig:ellips}}
\end{figure*}

Relationship \eqref{eq:q22} is depicted in Figure~\ref{fig:normC}; for a typical compactness, $\mathcal{C} \approx 0.17$, we have a $\approx 30\%$ increase in $Q^{22}_{0}$ relative to the Newtonian expression. This occurs essentially because the absolute magnetic energy also increases as a function of compactness implying that, if $B_{0}$ is fixed, the quadrupole moment should be larger. Despite this increase, as the spindown luminosity is also boosted by relativistic factors \citep{rez04}, a given period ($P$) and period derivative ($\dot{P}$) measurement corresponds to a weaker $B_{0}$ inference in GR (see Sec.~\ref{sec:detect}). Thus, the inferred magnetospheric energy is actually likely to be \emph{lower}, compared to the Newtonain case, from a given set of timing parameters. For $\mathcal{C} \approx 0.17$ the untwisted energy is $\approx 45\%$ of the Newtonian value for fixed $P \dot{P}$ (see figure 6 in Ref.~\cite{stef25}) meaning that the untwisted quadrupole moment is lower by $\approx 40\%$ (accounting for the increase seen in Fig.~\ref{fig:normC}). Although this is a non-negligible change, with the exception of recycled pulsars or objects born from merger events we do not expect much variation in $\mathcal{C}$ between members of the isolated neutron-star population.

\subsection{Comparison with hydromagnetic deformations} \label{sec:hydromag}

Many studies of equilibria in the interior of neutron stars stipulate a larger quadrupole moment than that of the magnetosphere \eqref{eq:q22}.  
{The characteristic strain associated with the interior follows that of expression \eqref{eq:h0} except with the fluid quadrupole moment \cite{aasi14},
\begin{equation}
    h_{0}^{\rm fluid} = \frac{16 \pi^2 G}{c^4} \frac{|Q^{22}_{\rm fluid}| f^2}{d}.
\end{equation}
Notably, X-ray \cite{mak14,mak21} and radio \cite{des24} modulations have been interpreted as evidence for precession and large quadrupole moments in some magnetars. 
For example, high-cadence radio monitoring of the magnetar XTE J1810--197 following its outburst in 2018 led \citet{des24} to infer a quadrupole moment of order $|Q^{22}_{\rm fluid}| \gtrsim 10^{38} \text{ g cm}^2$. 
Given the (GR-corrected) polar field strength estimate of this object of $B_{p} \sim 10^{14}$~G \cite{stef25}, we may anticipate a ratio $|Q^{22}_{\rm magneto}|/|Q^{22}_{\rm fluid}| \lesssim 10^{-2}$ from expression \eqref{eq:q22}.}

Even if the magnetospheric contribution is comparatively weak however, its detectability is largely independent of the interior. 
This is because its polarisation pattern (i.e., the phase portraits of $h_+$ and $h_{\times}$) is set by the distribution of surface currents \cite{naz20,cont24,kouv24}, while that of the interior is determined by the equation of state, poloidal-toroidal partition, and other factors \cite{lm13,aasi14}. 
For example, in the simple case of a biaxial deformation and ignoring detector-frame corrections, the plus polarisation profile of the total system reads \cite{naz20}
\begin{equation} \label{eq:hplus}
\begin{aligned} 
h_+ \approx& h^{\rm magneto}_{0}\sin\vartheta_0\Big[\sin\vartheta_0\left(1+\cos^2\iota\right)\cos \left(2\Omega \tau+2\varphi_0\right)\\
&+\cos\vartheta_0\cos\iota\sin\iota\sin \left(\Omega \tau+\varphi_0\right)\Big]\\
&+ \frac{1}{2} h^{\rm fluid}_{0}\left(1+\cos^2\iota\right)\cos \left(2\Omega \tau\right),
\end{aligned}
\end{equation}
where $\iota$ is the inclination from the rotation axis to the line-of-sight, $\Omega = 2\pi/P$ is the angular frequency, $\tau$ is the retarded time, and $\vartheta_0$ and $\varphi_0$ respectively denote the angles between the Cartesian $z$- and $x$-axes with the magnetic axis in the rotating frame.
An important feature evident in expression \eqref{eq:hplus} is that the phase offset $\varphi_{0}$ is present only in the 1PN waveform, $h_{0}^{\rm magneto}(t)$. 
As explained by \citet{naz20}, it arises because the PN phase portrait is set by gradients of the magnetic pressure alone, while the hydromagnetic term depends on the total Lorentz force. 
The distinction becomes even more apparent for tangled fields in the interior, as explored in detail by \citet{lm13}, where the star may also radiate at other harmonics of $\Omega \tau$ depending on the Euler angles. 

Over the course of a multi-year observational campaign with DECIGO, the superposition of terms within \eqref{eq:hplus} and $h_{\times}(t)$ could therefore be independently visible in the full waveform if $\varphi_0 \neq 0$. 
{As the detectability of $h_{0}^{\rm fluid}$ has been explored extensively in the literature (see, e.g., Refs.~\cite{gg18,col08,c09,lm13,cr13,mast11,mast15,hask08,fr12,dex19,lan13,sed25}), we henceforth ignore it and consider only the detectability of the (independent) component $h_{0}^{\rm magneto}$ defined by expression \eqref{eq:h0}.}
Although we focus on the amplitude \eqref{eq:h0} in this paper to provide sky- and polarisation-averaged estimates (Sec.~\ref{sec:detect}), the above should be kept in mind since, even if the hydromagnetic deformation dominates, the magnetospheric signal will not be subsumed in general. 
In fact, a detection of both $h_{0}^{\rm fluid}(t)$ and $h_{0}^{\rm magneto}(t)$ could be used to independently constrain the poloidal-toroidal partition and near-surface field strength, which would provide more information than the detection of an individual piece \cite{msm15}.

\section{Results for twisted models} \label{sec:magmodels}

Here we consider a set of models, varying the main parameters described in Sec.~\ref{sec:numerics}. To highlight the plausible extrema, we consider a sequence of models of compactness $\mathcal{C} \leq 0.25$.

Figure~\ref{fig:ellips} shows quadrupole moments \eqref{eq:massquadnaxi} for fixed $\mathcal{C} = 0.17$ computed from numerical outputs of twisted magnetic fields for varying $\alpha_{0}$ (left), $\sigma$ (middle), and $\theta_{1}$ (right), relative to the untwisted $(\alpha_{0} = 0)$ state. 
For reference, when $\alpha_{0}$ is held fixed, a base value of $\alpha_{0} = 1.5 R^{-1}$ is taken. The other reference values are $\theta_{1} = \pi/4$ and $\sigma = 0.2$, as appropriate. The endpoints of each sequence are set by demanding equality between energy integrals computed using either the virial theorem or the field directly. In particular, if equality is not met within a numerically-acceptable tolerance this indicates a breakdown of the force-free assumption and thus solution non-existence; see Appendix B of SSP25 for details.

As expected, the greatest departure in $Q^{22}$, relative to the untwisted state, occurs for the largest values of $\alpha_{0} = 2.5 R^{-1}$ ($Q^{22}/Q^{22}_{0} \approx 0.97$), as this sets the overall scale for the twist. Importantly, because of the maximum energy-excess aspect of the magnetosphere problem highlighted earlier, only small adjustments (of order several percent) are found for any given parameter choices. Quadrupole shifts are more prominent for spot latitudes closer to the polar cap $(\theta \sim 0)$ due to the angular harmonic that weights the integrand in expression \eqref{eq:massquadnaxi}, as indicated by the final panel of Fig.~\ref{fig:ellips}, though still the impact is small because the solution-space for $\alpha_{0}$ becomes more restrictive. We see that the ratio dips below unity in twisted cases, indicating that toroidal pressures slightly \emph{reduce} the GW luminosity. This phenomenon is similar to that observed for hydromagnetic deformations in the interior, where poloidal/toroidal fields induce competing oblate/prolate deformations \cite{col08,hask08,c09}. In the magnetospheric case, however, since the free energy of \emph{stable} cases must be relatively small for any combination of parameters, the quadrupole moment cannot flip sign. 

\begin{figure}
    \includegraphics[width=0.49\textwidth]{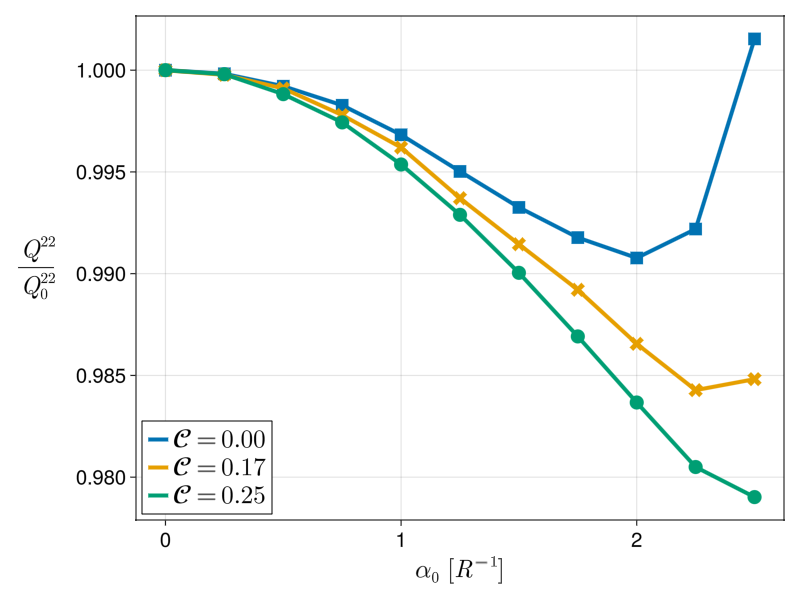}
    \caption{Mass quadrupole moments of twisted GR magnetospheres, normalised by the current-free value \eqref{eq:q22}, as a function of $\alpha_0$ for $\mathcal{C} = 0$ (blue squares), $\mathcal{C} = 0.17$ (yellow stars), or $\mathcal{C} = 0.25$ (green circles).}
    \label{fig:grquads}
\end{figure}

Fixing $\theta_1 = \pi/4$ and $\sigma = 0.2$, quadrupole moments for sequences of varying $\alpha_0$ and $\mathcal{C}$ are shown in Figure~\ref{fig:grquads}, which can be compared with the left-hand panel of Fig.~\ref{fig:ellips}. Graphical representations for some of these twist profiles are shown in figure 4 in SSP25. Again we see that, even for extreme twists $(\alpha_0 \sim 2.5 \Rs^{-1})$ where the force-free approximation is borderline violated, $Q^{22}$ is smaller than expression \eqref{eq:q22} by at most $\approx 3\%$. For ultra-compact stars with $\mathcal{C} = 0.25$ the departure is greatest, while for Newtonian stars we see an uptick because toroidal pressures become more dilute (cf. figures 4d--f in SSP25). The smallness of these changes demonstrate that, regardless of compactness or twist, we expect \emph{universality} in the sense that the magnetospheric quadrupole moment is roughly constant for fixed values of $B_{0}^2 R^5$ and $\mathcal{C}$.

Although we have not presented an exhaustive sweep, the conclusions are robust: the maximum free energy that a \emph{stable} magnetosphere can hold is at most of a few tens of percent independently of the stellar compactness. This excess applies to axisymmetric cases, where we find $Q^{22}_{0}/Q^{22} \approx 1.2$ at the limit of the stable branch (see figure 3 in SSP25). For the arguably more realistic cases we have focussed on here with non-axisymmetric, localised spots, the energy discrepancy is at most of order $\approx 5 \%$. In general, increases in $Q^{22}$ correlate with the energy excess, as is clear from expression \eqref{eq:massquadnaxi}. For example, the $\alpha_{0} = 2.5 R^{-1}$ case has an excess of $\sim 3\%$ and the quadrupole moment decreases by approximately the same factor. We may therefore anticipate, at most, a factor $\lesssim 1.2$ enhancement for a (likely unrealistic) case that is in some sense maximally twisted with respect to all parameter choices (cf. figure 3 in SSP25). This indicates that GWs emitted by twisted magnetospheres are practically indistinguishable from untwisted ones, and an eventual detection could therefore be used to directly probe the combination $B_{0}^2 \Rs^{5}$ appearing in expression \eqref{eq:q22}.

\subsection{Dipole-plus-quadrupole fields} \label{sec:dipquad}

\begin{figure}
    \includegraphics[width=0.49\textwidth]{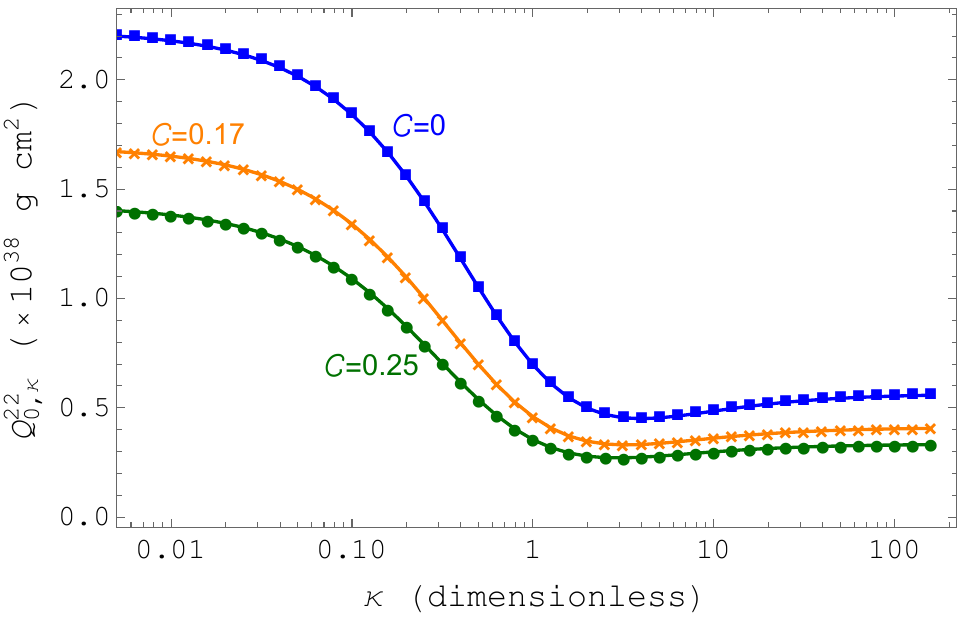}
    \caption{Variation of the (absolute value of the) quadrupole moment \eqref{eq:quadquadmom} for untwisted, dipole-plus-quadrupoles with $\mathcal{C} = 0$ (blue squares), $\mathcal{C} = 0.17$ (orange crosses), or $\mathcal{C} = 0.25$ (green circles) where we fix $\Rs =10$~km and the surface field strength maximum to $\tilde{B} = 2 \times 10^{15}$~G. 
    \label{fig:normQ}}
\end{figure}

One complication with the above picture is that of multipolarity. Indeed, there is evidence that the near-surface fields of (at least some) magnetars may contain significant, non-dipole components (see below). In this section, we explore how \emph{magnetic} quadrupoles may adjust the anticipated signal.

The untwisted quadrupole moment \eqref{eq:massquadnaxi}, to leading order in $\mathcal{C}$, reads (for $R_{\rm LC} \gg \Rs$)
\begin{equation} \label{eq:quadquadmom}
Q^{22}_{0,\kappa} =  -\frac{B_{0}^2  \Rs^5\left(63 + 16 \kappa^2 +105 \mathcal{C} +60 \mathcal{C}  \kappa^2 \right)}{315 c^2} + \mathcal{O}(\mathcal{C}^2),
\end{equation}
where we remind the reader that $\kappa$ is defined as the relative
strength between the magnetic quadrupole and dipole components.
Note that expression \eqref{eq:quadquadmom} grows indefinitely with increasing $\kappa$ because the field strength itself grows with $\kappa$. If instead we fix the surface field-strength maximum to a value $\tilde{B}$ (say), then we have $B_{0} \approx \tilde{B}/(1+ \kappa)$ depending on the compactness and thus, for quadrupole-dominated systems, $Q^{22}_{0,\kappa \to \infty}/Q^{22}_{0,0} \approx 0.3$. The intermediate regime is instead shown in Figure~\ref{fig:normQ}, indicating a local minimum around $\kappa \sim 3$ and that for $\kappa >0$ we have a decrease in the mass-quadrupole moment for fixed surface-maximum $\tilde{B}$. The absolute difference as a function of $\kappa$ is smaller as $\mathcal{C}$ increases because $B^2$ becomes more concentrated near the surface (see expression~\ref{eq:psifn}) and, having fixed $\tilde{B}$, larger radii contribute less to the integral. The overall value of $Q^{22}_{0,\kappa}$ is lower for more compact stars for the same reason.

\begin{figure}
    \includegraphics[width=0.45\textwidth]{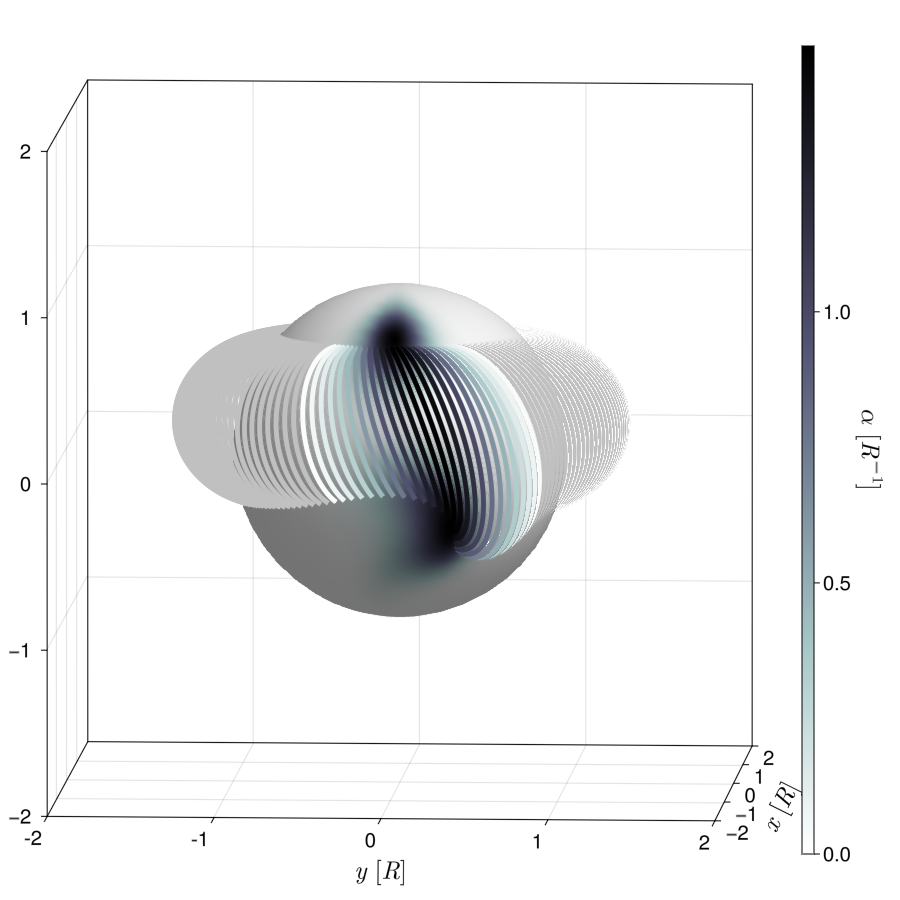} \\
    \includegraphics[width=0.45\textwidth]{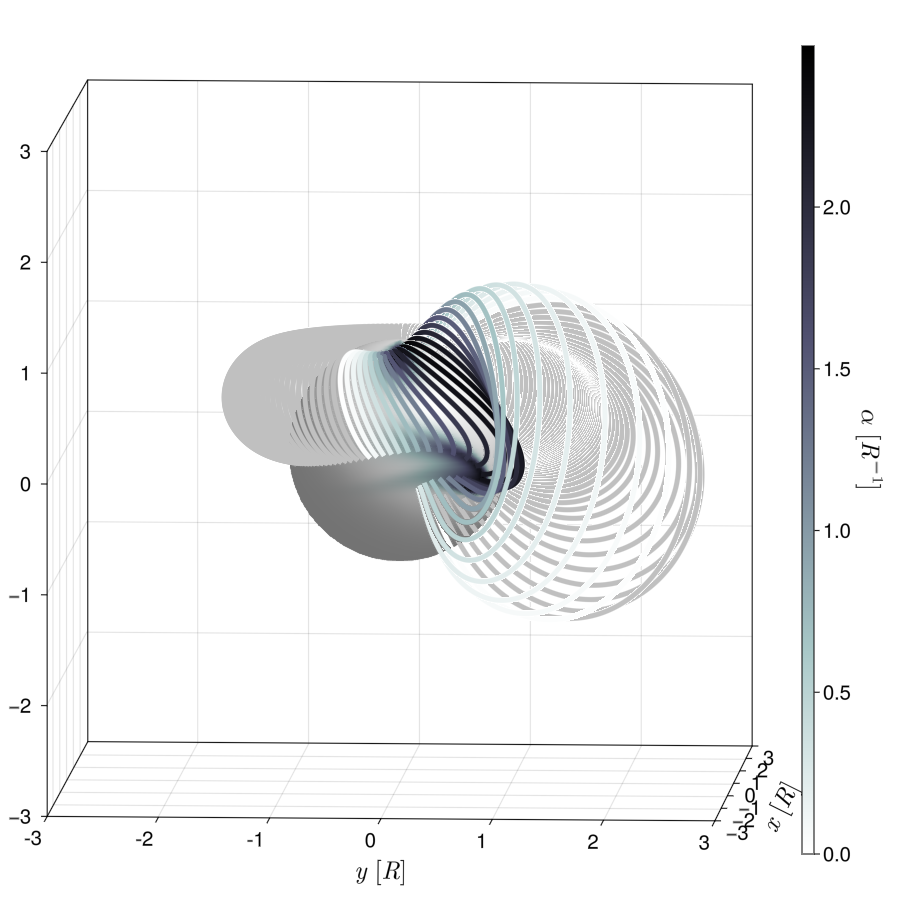}
    \caption{Similar to the top panel of Fig.~\ref{fig:dipolelines} but for dipole-plus-quadrupole fields with $\mathcal{C}=0.17$,  $\alpha_{0} = 2.5 R^{-1}$, $\theta_{1} = \pi/6$, $\sigma = 0.2$ and $\kappa = 0.5$ (top) or $\kappa = 2$, (bottom). 
    \label{fig:quadlines}}
\end{figure}

We present a solution to the constrained system \eqref{eq:magnetoeqn} and \eqref{eq:alpha_constraint} for a dipole-plus-quadrupole field with twist to compare with the above results. One particular realisation with $\kappa = 0.5$, but otherwise identical parameters to that shown in Fig.~\ref{fig:dipolelines}, is shown in the top panel of Figure~\ref{fig:quadlines}. We find in this case that $Q^{22}$ increases by $\approx 3.5\%$ relative to \eqref{eq:quadquadmom}. Notably, the overall value is lower than the untwisted dipole configuration by a factor $\sim 0.8$ if the surface field value is held fixed (see Fig.~\ref{fig:normQ}).

Cyclotron lines from SGR 0418+5729 point towards a near-surface field of $\gtrsim 10^{14}$~G if interpreted as a proton resonance while the inferred dipolar field is $\sim \text{few} \times 10^{12}$~G \citep{tiengo13}. If indeed $\tilde{B} \gtrsim 10^{14}$~G we should have $\kappa \gtrsim 10$ if a quadrupole field is dominant, so that $Q^{22}_{0, \kappa}$ may be weaker by a factor $\sim 4$ (Fig.~\ref{fig:normQ}) relative to the dipole case for this object or other `low-field' sources like Swift J1822.3--1606 \citep{cast16}. The magnetosphere is also likely rather twisted in such cases as these stars exhibit outbursts despite their `low' dipole moments. On the other hand, for $\kappa = 2$, $\alpha_{0} = 2.5 R^{-1}$, $\theta_{1} = \pi/6$, and $\sigma = 0.2$  -- depicted in the bottom panel of Fig~~\ref{fig:quadlines} -- we find a change over expression \eqref{eq:quadquadmom} that is at a sub-percent level. As $\kappa$ increases, the relative change appears even lower. That is, strong quadrupolar components actually reduce the disparity in $Q^{22}$ between twisted and potential states and \emph{promote universality}, especially for compact stars.

Overall, therefore, the conclusions discussed previously still hold for multipolar fields with the important distinction being that a detection would not probe the dipole component of the field but rather the near-surface maximum, $\tilde{B}$, within some tolerance (Fig.~\ref{fig:normQ}). Although only shown here for dipole-plus-quadrupole fields, twists cannot significantly increase $Q^{22}$ for any multipolarity because there is always an energy-excess maximum that is fairly restrictive (see references in Sec.~\ref{sec:maxtwist}). For any given field structure, the integrand will be dominated by contributions near the surface maximum and thus $Q^{22}$ cannot vary significantly, meaning that tight constraints on $\tilde{B}$ can be placed if the source distance is known and the radius is approximately $12$~km (as inferred from multi wavelength data; see figure 2 in Ref.~\cite{skk24}).

\section{Implications for detectability} \label{sec:detect}

In order to quantify the detectability of magnetospheric GWs, we can estimate the SNR. Tracking an object via matched-filtering, from GW frequencies $f_{i}$ until $f_{f}$ (twice that of the spin frequency), the SNR reads \cite{dall09}
\begin{equation} \label{eq:snrform}
\text{SNR} = 2 \left[ \int^{f_{f}}_{f_{i}} df \frac{|\tilde{h}(f)|^2}{S_{n}(f)} \right]^{1/2}, 
\end{equation}
where $\tilde{h}(f)$ is the Fourier transform of $h(t)$ defined through
\begin{equation}
\begin{aligned}
    h(t) =& h_{0} \Big[ \frac{1}{2} F_{+}(t,\psi) \left(1 + \cos^2 \iota\right) \cos \tilde{\phi}(t) \\
    &+ F_{\times}(t,\psi) \cos \iota \sin \tilde{\phi}(t) \Big],
    \end{aligned}
\end{equation}
which, as noted in Sec.~\ref{sec:hydromag}, depends on the antenna-pattern response functions for the two polarisation states ($F_{+}$ and $F_{\times}$), polarisation angle $\psi$, and phase evolution $\tilde{\phi}(t)$ (see Ref.~\cite{aasi14} for details). In expression \eqref{eq:snrform}, $S_{n}$ denotes the (instrument-dependent) noise spectral density of the detector. {We emphasise again that we discuss only the detectability of the magnetospheric contribution (i.e., setting $h_{0} = h_{0}^{\rm magneto}$), which varies little between models due to the universal behaviour explored throughout.
To estimate the detectability of the \emph{total} system, one should substitute a combined amplitude $h(t) \equiv \sum_{i} \left(F_{+} h^{i}_{+} + F_{\times} h^{i}_{\times}\right)$ \cite{thorne95} into the Fourier integral \eqref{eq:snrform} (see, e.g., Ref.~\cite{msm15} for a discussion).
}

Although several studies in the literature consider newborn magnetars with $f_{i} \sim 1$~kHz, phenomena related to the supernovae and magnetar wind, together with difficulty in tracking the spin to maintain phase coherence, imply that \eqref{eq:snrform} is likely to provide an overestimate in such cases \citep{lj20}. Considering a more conservative scenario where the star is observed starting from a modest frequency $f_{i} \sim 10$~Hz until $f_{f} \sim 1$~Hz -- plausible if the source pulses in the X- or radio bands so that timing is possible -- we estimate the SNR \eqref{eq:snrform} using the noise curves from figure 2 of \citet{yagi13} where $S_{n} \approx 10^{-49} (f/1\text{ Hz})^{2} \text{ Hz}^{-1}$ around $f \gtrsim 1$~Hz for DECIGO (see the discussion on curve fitting in Ref.~\cite{sp25}).

\subsection{Dipole fields} \label{sec:dipsnr}

To integrate \eqref{eq:snrform}, we need to assume a spindown law to determine the frequency evolution. The GW power reads \cite{dall09}
\begin{equation} \label{eq:gwpower}
\begin{aligned}
    P_{\rm GW} &= \frac{2^{11} \pi^6 G |Q^{22}|^2}{5 c^5 P^6} \\
    &\approx 10^{22} \left(\frac{Q^{22}} {10^{38} \text{ g cm}^{2} }\right)^{2} \left(\frac{1 \text{ s}}{P}\right)^{6} \text{ erg s}^{-1},
    \end{aligned}
\end{equation}
which is always small relative to electromagnetic spindown unless the object has a millisecond spin period {and the fluid quadrupole is very large}, meaning it can be ignored for a slow rotator. For an inclined star with $P \gg 1 \text{ ms}$ surrounded by magnetospheric plasma, we thus have \citep{rez04,petri16},
\begin{equation} \label{eq:sd}
P \dot{P} \approx    \frac {2 \pi^2 B_{p}^2 \Rs^6 \left(a + b \sin^2 \chi \right)}{3c^3\left(1 - 2 \mathcal{C} \right)^{7/2} I_{0}},
\end{equation}
for (Tolman-VII) moment of inertia $I_{0} \approx 2 M R^2 (1 - 1.1 \mathcal{C} - 0.6 \mathcal{C}^2)^{-1}/7$, inclination angle $\chi$, and polar-dipole strength $B_{p} \gtrsim 2 B_{0}$. We consider $a=1$ and $b=1.5$ as in SSP25. Expression \eqref{eq:snrform} can be integrated straighforwardly, averaging over the antennae response functions $F_{+,\times}$ \citep{dall09}. Doing so for\footnote{Note that a signal accumulation time of $\sim 2$~yr is needed for $f_{i}$ to reach $f_{f}$ if $B_{p} \sim 2 \times 10^{15}$~G.} $f_{i} = 10 \text{ Hz} = 10f_{f}$ and assuming $\chi = \pi/2$ for optimal GW losses we have, for the untwisted dipole \eqref{eq:puredip},
\begin{equation} \label{eq:snrbase}
    \text{SNR} \approx 0.1 \left(\frac{B_{0}}{10^{15} 
 \text{G}}\right)^{2} \left(\frac{\Rs}{12 \text{km}}\right)^{3} \left(\frac{M}{1.4 M_{\odot}}\right)\left(\frac{\text{kpc}}{ d}\right).
\end{equation}

For a strongly twisted case with $\alpha_{0} \approx 2.5 R^{-1}$, both the quadrupole moment and SNR change by at most a few percent (see Fig.~\ref{fig:ellips}). For an object like SGR 1806--20 with $B_{0} \sim 10^{15}$~G, marginal detections out to the nearby ($\gtrsim$~100 pc) star-forming complex Sco OB2 are plausible for dipolar fields. Note, however, that in cases with twist, expression \eqref{eq:sd} may no longer be appropriate: the magnetospheric `Y-point' (i.e. where the magnetospheric current sheet meets the separatrix between open and closed field lines) should migrate towards the stellar surface as the twist increases. Physically this occurs as toroidal pressures inflate the field lines (Fig.~\ref{fig:dipolelines}), forcing more of them through the light cylinder as the twist grows. Some degree of caution should be applied therefore since, if the star spins down faster due to excess twist, there is less time to accumulate signal when observing within fixed frequency bins (see figure 5 in Ref.~\cite{ng25}). This also applies to more compact stars, which have an enhanced spindown luminosity for fixed $B_{0}$ due to equation \eqref{eq:sd}: the `Newtonian SNR' for the same parameters is $\approx 0.26$. Twist is actually detrimental to detection therefore, even if $Q^{22}$ itself hardly changes, especially due to achromatic timing noise from torque variability \cite{suva16}. 

\subsection{Dipole-plus-quadrupole fields} \label{sec:dipquadsnr}

\begin{figure}
    \includegraphics[width=0.49\textwidth]{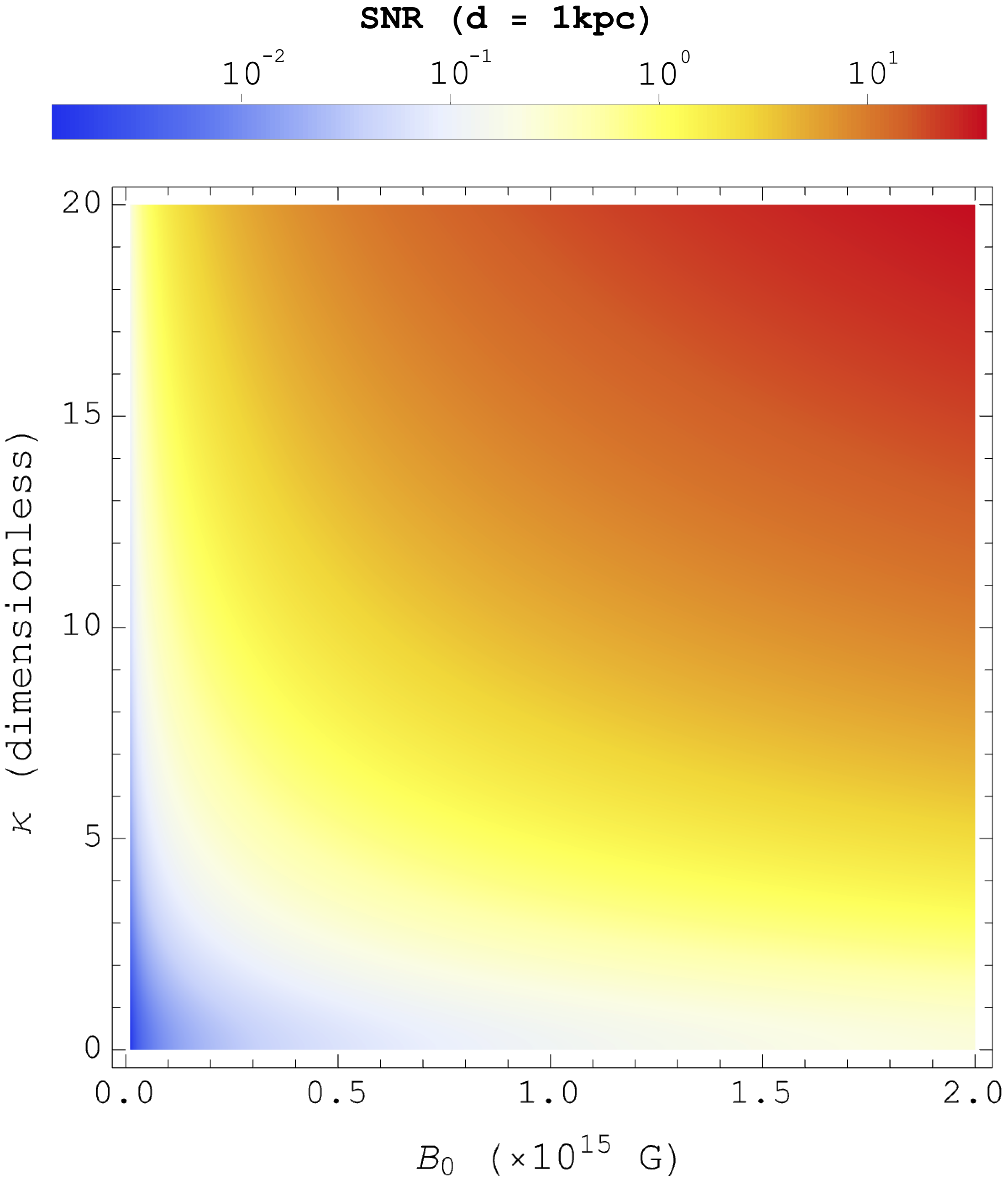}
    \caption{Values of the expected SNR \eqref{eq:snrform} for untwisted magnetospheres as a function of $B_{0}$ (horizontal) and $\kappa$ (vertical) for compactness $\mathcal{C} = 0.17$; redder shades indicate greater detectability odds.
    \label{fig:snrdipquad}}
\end{figure}

Although the quadrupole moment does not change much between dipolar and dipole-plus-quadrupole cases, the detectability of the source may be impacted significantly. This is because even if the quadrupole moment \eqref{eq:quadquadmom} is reduced, spindown is stalled for fixed $\tilde{B}$ (with a braking index $n \approx 5$, as for GW-induced spindown, for large $\kappa$) meaning that signal may be accumulated over longer time spans. This is essentially the opposite effect to that described above for twist-assisted spindown. 

Figure~\ref{fig:snrdipquad} depicts the SNR \eqref{eq:snrform} for dipole-plus-quadrupole fields as functions of $B_{0}$ and $\kappa$, where the Fourier-space integration window is the same as considered previously ($f_{i} = 10 f_{f} = 10$~Hz). Here, we imagine that a surface dipole component has been independently measured (e.g. from pulse timing) while the true near-surface field is unknown. In general, larger values of $\kappa$ and $B_{0}$ lead to greater SNR because both parameters boost the effective quadrupole moment \eqref{eq:massquadnaxi}. Note in particular that, even for dipolar strengths of $B_{0} \sim 10^{15}$~G, SNRs of $\sim 10$ for an object at $d \sim 1$~kpc could theoretically be achieved if the system hosted a surface component with $\tilde{B} \gtrsim 10^{16}$~G because spindown is hardly accelerated at spin periods of $f \sim 1$~Hz by multipoles as the light-cylinder lies very far from the surface. Even given some universal behaviour for the quadrupole moment itself, detection prospects may thus improve significantly for multipolar configurations. (Though caution should be applied for low $B_{0}$ since the mission lifetime, estimated to be at least three years \cite{kaw21}, sets a cutoff to the Fourier integration window).

\subsection{Nearby magnetars?} \label{sec:nearby}

We have argued that detections out to $\lesssim 1$~kpc may be feasible for magnetars entering the DECIGO band with $B_{0} \gtrsim 5 \times 10^{14}$~G and $\kappa \gtrsim \text{few}$ (see Fig.~\ref{fig:snrdipquad}). It is natural to question therefore whether one expects any such objects within our neighbourhood. A reasonable upper limit of $\sim$~few per century may apply for the Galactic birth rate of objects with comparable fields \citep{gull15,beni19}. Taking a rough Milky Way volume of $\sim 10^{3} \text{ kpc}^3$, we could thus expect an (upper limit) birth rate of $\lesssim 10^{-4}$ for such objects, per year, within $\sim 1$~kpc of Earth. 

Taking equation \eqref{eq:sd}, such an object could reside within the peak of the DECIGO sensitivity bowl for $\lesssim 10^{3}$~yr for a birth period of $\sim 10$~ms. Given an (optimistic) mission lifetime of $\sim 10$~yr \citep{kaw21}, we can thus anticipate that $\lesssim \mathcal{O}(1)$ sources may be detected, noting that our SNR estimates \eqref{eq:snrform} assume matched filtering and thus accurate observations of $P(t)$, even ignoring complications associated with tracking in the presence of achromatic timing noise \cite{suva16}. Nevertheless, because (i) nearby star-forming clusters may host overdensities of magnetars, (ii) statistics based on small sample sizes may be subject to revision in future when deeper observations are available, and (iii) some evolutionary scenarios predict either field reemergence \citep{sl12,lander24} or long-term dormancy \citep{torr16,sdp25}, some optimism is warranted.

\citet{pop03} have argued in fact that there is a local excess of neutron stars surrounding the solar system owing to the proximity of the Gould Belt and the OB-star complexes it houses. This is consistent with observations of the `magnificient seven' (M7), as it is otherwise difficult to explain why so many neutron stars reside within several hundred parsecs of Earth. \citet{motch05} used optical-image, Subaru data to constrain the proper motion of RX J1605.3+3249 and other members of the M7, finding apparent trajectories consistent with a birth in a nearby Sco OB2 complex. An overdensity of local magnetars -- which may appear X-ray dim in old age -- is therefore plausible. This conclusion is supported by the recent discovery of CHIME J0630+25, a long-period pulsar ($P \approx 7$~min \cite{dong24}) whose radio timing and dispersion measure indicate $B_{0} \lesssim 10^{15}$~G and $d \sim 100$~pc (assuming the object is a neutron star; see Ref.~\cite{sdp25} for a discussion). 

\section{Discussion} \label{sec:discussion}

In this paper we explore how currents impact on the detectability of GWs from magnetar magnetospheres. Because solutions to equations \eqref{eq:magnetoeqn} and \eqref{eq:alpha_constraint} can sustain a \emph{maximum} free energy relative to an untwisted state of only $\approx 30\%$ if there are no magnetic islands \citep{stef25}, GW amplitudes associated with the magnetosphere are confined to a narrow range. The strain \eqref{eq:h0} is controlled by the combination of $B_{0}^2 R^{5}$ (expression~\ref{eq:q22}) regardless of the twist particulars (see Fig.~\ref{fig:ellips}). It is in this sense we argue for a universal behaviour between objects, akin to others described in the literature for various aspects of the interior \citep{yy13}. Although detection prospects are marginal for Galactic objects even with the planned, space-based detector DECIGO -- most sensitive in the $\lesssim 10$~Hz band -- universality is of theoretical interest as probing the near-surface fields of magnetars is difficult owing to timing anomalies, state switching, and other phenomena \cite{mel99,low25}. 
As emphasised in Sec.~\ref{sec:hydromag}, the magnetospheric signal sets an effective floor to the GW luminosity and thus our results should be thought of as quantifying detectability in the most conservative scenario.
Nevertheless, since the phase portrait is different between the hydromagnetic and magnetospheric pieces (equation \ref{eq:hplus}), the detectability of the latter is largely independent of the former.

With respect to multimessenger connections, suppose that the neutron star undergoes some (sub-)crustal event such that energy is deposited into the magnetosphere. This energy will inject helicity, and thus may lead to a modulation of the characteristic strain until the system relaxes over a diffusion timescale which could reach $\sim 40$~yr for moderate twists in an X-ray dim magnetar with luminosity $\sim 10^{33} \text{ erg s}^{-1}$ (see expression 38 in Ref.~\cite{thom02}). The quadrupole moment may therefore `wander' in a way that can be tied to timing anomalies. Such modulations could be calculated using a set of models, like that we have constructed here, that are sequenced by assuming that the twist decays linearly -- as is roughly consistent with post-outburst observations  \cite{mpm15,low23,low25}. In principle, signals with sufficient SNR may be disentangled to reveal such features coincident with X-ray activity (though cf. Ref.~\cite{zink12}). Next-generation technology could thus provide detailed information about the ignition mechanism of flares (i.e., whether they are magnetospheric in origin or crustal) and spindown for nearby ($\sim 100$~pc) objects like CHIME J0630 \cite{dong24}. 

Another possibility concerns a stochastic GW background. If indeed all neutron star magnetospheres produce a characteristic signal that permeates the universe, this may be searched for using pulsar-timing arrays \citep{jenet05}. If detectable, this could constrain the distribution of near-surface magnetic field strengths. This, and other aspects described above, will be investigated elsewhere.

We close this paper with some general remarks on the validity of our approach. In particular, explicit time-dependence in the quadrupole moment (sourced, in our case, by the magnetic field directly) is necessary to drive GW emissions. The fact that we solve for the magnetospheric structure via the static equations \eqref{eq:magnetoeqn} and \eqref{eq:alpha_constraint} is, therefore, not really consistent with our main motivation. Nevertheless, deducing the biaxial (magnetic) ellipticity is a common practice in the literature \cite{hask08,c09,col08,mast15} and is expected to qualitatively reflect GW predictions for stars computed self-consistently with a misaligned inclination angle (though see also Ref.~\cite{lm13}). In the case of the fluid interior, this is typically justified because advective terms are small for low rotation rates (though cf. Ref.~\cite{suvg23}). For the magnetosphere, solutions to the so-called `pulsar equation' differ qualitatively from their static counterparts. Even so, because arguments for the maximal twist based on the stability of solutions to equation \eqref{eq:magnetoeqn} should hold for rotating systems \citep{uz02,ng25}, we still expect `universality'.

\section*{Acknowledgements}
We thank Juan A. Miralles for discussions.
AGS acknowledges funding from the European Union's Horizon MSCA-2022 research and innovation programme ``EinsteinWaves'' under grant agreement No. 101131233. 
We acknowledge support from the Prometeo excellence programme grant
CIPROM/2022/13 funded by the Generalitat Valenciana. We also acknowledge 
support through the grant PID2021-127495NB-I00 funded by MCIN/AEI/10.13039/501100011033 and by the European Union, as well as the Astrophysics and High Energy Physics programme of the Generalitat Valenciana ASFAE/2022/026 funded by MCIN and the European Union NextGenerationEU (PRTR-C17.I1).

\end{document}